\documentstyle[preprint,prd,aps,floats,epsf,tighten]{revtex}
\begin{document}

\preprint{\vbox{
\hbox{FSU-SCRI-99-68}
\hbox{JLAB-THY-99-33}
\hbox{UCD 1999-20}
\hbox{hep-lat/9910041}
}}

\title{Chiral condensate in the deconfined phase of
quenched gauge theories}
\author{Robert G. Edwards}
\address{
SCRI, Florida State University, 
Tallahassee, FL 32306-4130, USA}
\address{
Jefferson Lab,
12000 Jefferson Avenue,
MS 12H2,
Newport News, VA 23606, USA}
\author{Urs M. Heller}
\address{
SCRI, Florida State University, 
Tallahassee, FL 32306-4130, USA}
\author{ Joe Kiskis}
\address{ Dept. of Physics, University of California,
Davis, CA 95616}
\author{ Rajamani Narayanan}
\address{ American Physical Society, One Research Road, Ridge, NY 11961 }
\maketitle 

\begin{abstract}
We compute the low lying spectrum of the overlap Dirac operator in the
deconfined phase of finite-temperature quenched gauge theory.
It suggests the existence of a chiral condensate which we confirm with
a direct stochastic estimate.  We show that the part of
the spectrum responsible for the chiral condensate can be understood
as arising from a dilute gas of instantons and anti-instantons.
\end{abstract}

\section {Introduction}

The fermion spectrum near zero eigenvalue is closely related to gauge
field topology and to chiral symmetry breaking. While there is
considerable experimental and theoretical support for chiral symmetry
breaking in gauge theories with dynamical quarks at zero temperature
and theoretical arguments for its restoration above a critical
temperature,
the situation is less clear for the nominally simpler
case of fermions in the background of quenched gauge fields. To
improve our understanding of the quenched, deconfined phase, we have
studied the spectrum of the hermitian overlap Dirac
operator~\cite{Herbert} in that region. We find a segment of the
spectrum concentrated at and near zero eigenvalue and separated from
the bulk of the spectrum. The bulk of the spectrum begins to rise
rapidly at larger eigenvalues. The exactly zero eigenvalues are
associated with the global topology of the gauge field configurations.
The statistical properties of the small eigenvalues are in
correspondence with predictions from a dilute gas of instantons and
anti-instantons. Small, non-zero eigenvalues with these properties
give rise to a finite chiral condensate.

Gauge field topology plays a central role in QCD.  The presence of
gauge field configurations with non-trivial topology indicates that
massless fermions will have exact zero modes. These zero modes cause
an explicit breaking of the axial U(1) symmetry and result in a
massive $\eta^\prime$~\cite{Hooft}. Conventional wisdom says that the
axial U(1) symmetry remains broken at all temperatures since one does
not expect a complete suppression of non-trivial gauge field
backgrounds. Recent studies using dynamical staggered fermions
indicate that the axial symmetry most likely remains broken at high
temperatures although the magnitude might be considerably smaller than
at low temperatures~\cite{Columbia,MILC_U1A,KLS}.
Since topology is also expected
to be suppressed at high temperatures~\cite{topology}, this result is
consistent with expectations.  Chiral symmetry breaking, on the other
hand, comes from the finite density of eigenvalues near
zero~\cite{Casher}, so we expect this density to be zero at high temperature
in full QCD since chiral symmetry is restored in that case.

The most likely scenario for the spectrum of the massless Dirac
operator at high temperatures is to have a delta function at zero due
to topology, followed by a gap and a continuous spectrum of
eigenvalues resulting in a theory with unbroken chiral symmetry and a
broken axial U(1) symmetry.  If we adopt the instanton picture for
topology, we expect a dilute gas of instantons and anti-instantons at
high temperature since the topological susceptibility is highly
suppressed. However, this dilute gas of instantons and anti-instantons
should not give rise to a chiral condensate in high-temperature, full
QCD. A natural explanation in the context of instanton models is that
instantons and anti-instantons form molecules at high
temperatures~\cite{Schafer}. One expects that this formation of
molecules is primarily due to interactions induced by
fermions~\cite{Schafer,molecule}.

In this paper, we will study the spectrum of the massless overlap Dirac
operator~\cite{Herbert} on the lattice in pure SU(2) and SU(3) gauge
theories on lattices with
$N_T=4$ in the deconfined phase. We will study several
different ensembles to understand the finite volume effects and also
the effect of going deeper into the deconfined phase. In addition to
directly studying the spectrum, we will also study the chiral
condensate and scalar susceptibility. In some ensembles, we will also
compare our results with those obtained using staggered fermions under
these conditions.

We have the following results:
\begin{itemize}
\item 
The overlap Dirac operator has exact zero eigenvalues indicating that gauge
field configurations with non-trivial topology persist in the high temperature
phase. 
\item
The spectrum of the
non-zero eigenvalues of the overlap Dirac operator has two parts separated by 
a region in eigenvalue where the density is essentially zero.
\begin{itemize}
\item The distribution of very small non-zero
eigenvalues of the overlap Dirac operator is well-described by a dilute gas 
of non-interacting instantons and anti-instantons.
\item The bulk part of the spectrum begins at larger eigenvalue and rises
smoothly and steeply.
\end{itemize}
\item 
The very small eigenvalues of the overlap Dirac operator have the properties
that are needed to give a chiral
condensate. The separation
between the very small eigenvalues and the bulk of the spectrum
results in a minimum in the scalar susceptibility.
\end{itemize}

These features are qualitatively different from the staggered Dirac
operator in a quenched theory.
The staggered Dirac operator does not have any exact zero
eigenvalues.  Studies with dynamical staggered fermions have indicated
that chiral symmetry is restored at high enough temperatures in QCD
with $N_f$ flavors of massless quarks and also in the quenched theory.
Early numerical studies of chiral symmetry restoration at high
temperatures using staggered fermions in the ``quenched''
approximation can be found in Ref.~\cite{Kogut}.  Further studies
using dynamical staggered fermions have established a phase transition
showing chiral symmetry restoration at high temperatures. Recent
reviews of the lattice results can be found in Ref.~\cite{FT_reviews}, and
Ref.~\cite{MILC} contains the most recent results for the $N_T=4$
phase transition.

The lack of exact zero eigenvalues in the staggered fermion spectrum
follows from the breaking of the continuum chiral and flavor
symmetry at finite lattice spacing~\cite{Vink}. The observed restoration
of chiral symmetry at high temperature is consistent with the prediction
of Ref.\cite{Tomboulis} for SU(2) gauge theory with staggered fermions.
However, a gap in the spectrum of the staggered Dirac operator at high
temperature has not been convincingly established, see {\it e.g.}
Ref.\cite{KLS}. A tail of small eigenvalues seems to persist, possibly
consisting of the ``shifted'' would be zero modes due to global topology.

The proof of chiral symmetry restoration at high temperature
in Ref.\cite{Tomboulis} is for dynamical
fermions, while we are only studying the quenched theory here. So, we could
explain the presence, and even accumulation, of very small eigenvalues in the
spectrum of the overlap Dirac operator as a
quenched artifact. This is not satisfactory, however, since staggered fermions
in a quenched theory show no such accumulation of small eigenvalues, but
rather a tapering tail of small eigenvalues, indicative of
a chirally symmetric phase. As is well known,
massless fermions are non-trivial to construct on the
lattice~\cite{Nielsen}. The overlap Dirac operator is one solution to
the problem. It is not ultra-local, {\it i.e.} its interaction size
is not finite, but its interactions are exponentially decreasing with
distance. It therefore does not fit into
the assumptions used in Ref.\cite{Tomboulis}, and the proof presented
there need not hold even in the dynamical theory with overlap
fermions. 
We do not address the
issue of the existence of a chirally symmetric phase in the dynamical
theory with overlap Dirac fermions in this paper.
                            
Care has to be taken in a finite temperature study in the quenched
approximation. In the pure gauge case, gauge field configurations with
the Polyakov loop phase near any $Z_N$ phase are equivalent, related by
a global $Z_N$ symmetry. However, the coupling to fermions is not $Z_N$
symmetric and the phase with the Polyakov loop along the positive
real axis is preferred. Configurations with ``non-trivial expectation value''
for Polyakov loop phase can give an ``unphysical'' signal for a chiral
condensate~\cite{Columbia} (negative for SU(2) and complex for SU(3)
with anti-periodic boundary conditions in the time direction for
fermions).  In pure SU(N) gauge theories, configurations with
expectation values for the Polyakov lines related by a simple $Z_N$
factor are equally likely. We will therefore generate pure gauge
ensembles so that every configuration has a positive expectation value
for the Polyakov loop and impose anti-periodic boundary conditions in
the time direction for fermions. So in a sense, this is not a strictly
quenched calculation, but has included in it by hand one important
feature of the theory with dynamical quarks.

The organization of the paper is as follows. In the second section,
we will briefly summarize what we need to know about the overlap Dirac
operator. 
We will present our results for the eigenvalue spectrum
in Section III and focus on the very small eigenvalues of the
overlap Dirac operator. We will show
that the spectrum of the very small non-zero and exact zero eigenvalues
is described quite well by a non-interacting dilute gas of instantons
and anti-instantons. We use the word ``dilute'' to emphasize that
topology is highly suppressed and that the spectrum of these very small
eigenvalues is separated from the bulk of the spectrum.
We will present our data for the chiral condensate and scalar susceptibility
and discuss the relationship between the small, non-zero eigenvalues and the
chiral condensate in Section IV. Our conclusions are in Section V.

Similar studies of the fate of chiral symmetry breaking in the deconfined
phase of quenched QCD have been done with domain wall fermions by the
Columbia group~\cite{DWF_CU} who measured $\langle \bar\psi\psi \rangle$
and found evidence of topology from its increase at small quark mass
and by Laga\"e and Sinclair~\cite{L_DKS} who studied low eigenvalues
and meson propagators and found evidence of topology from both.

\section{Overlap Dirac operator}
The massive overlap Dirac operator is given by~\cite{Herbert,Robert}
\begin{equation}
D(\mu) = {1\over 2}\left [ 1 + \mu + (1 -\mu ) \gamma_5 \epsilon(H_w)
\right ] 
\label{eq:overlap}
\end{equation}
with $0\le\mu\le 1$ describing fermions with positive mass all the way
from zero to infinity. The hermitian operator $H_w$ is just $\gamma_5D_w$
where $D_w$ is the usual Wilson-Dirac operator with a negative mass on
the lattice under consideration.

The propagator for external fermions is given by~\cite{Herbert,Robert}
\begin{equation}
{\tilde D}^{-1}(\mu) = (1-\mu)^{-1} \left[ D^{-1}(\mu) -1 \right]
\quad .
\label{eq:prop}
\end{equation}

In many cases, it is more convenient to use the hermitian version:
$H_o(\mu) = \gamma_5 D(\mu)$. It is easy to see that
\begin{equation}
H_o^2(\mu) = ( 1 - \mu^2 ) H_o^2(0) + \mu^2;\ \ \ \ 
[H_o^2(0), \gamma_5] = 0  .
\label{eq:chiral}
\end{equation}
Each eigenvalue $0 < \lambda^2 < 1$ of $H_o^2(0)$ is doubly degenerate
with opposite chirality eigenvectors. In this basis, $H_o(\mu)$ and $D(\mu)$
are block diagonal with $2 \times 2$ blocks, {\it e.g.,} $D(\mu)$:
\begin{equation}
\pmatrix{
         (1-\mu) \lambda^2 + \mu & (1-\mu) \lambda \sqrt{1-\lambda^2} \cr
         -(1-\mu) \lambda \sqrt{1-\lambda^2} & (1-\mu) \lambda^2 + \mu
  }
\end{equation}
where
\begin{equation}
\gamma_5 = \pmatrix{ 1 & 0 \cr 0 & -1 } .
\end{equation}

For a gauge field with topological charge $Q \ne 0$, there are, in addition,
$|Q|$ exact zero modes with chirality ${\rm sign}(Q)$ paired with
eigenvectors of opposite chirality and eigenvalue equal to unity.
These are also
eigenvectors of $H_o(\mu)$ and $D(\mu)$:
\begin{equation}
D(\mu)_{\rm zero~sector} : \quad \pmatrix{
                         \mu & 0 \cr 0 & 1 }
  \quad {\rm or} \quad
  \pmatrix{ 1 & 0 \cr 0 & \mu }
\end{equation}
depending on the sign of $Q$.

In the chiral eigenbasis of $H_o^2(0)$, the external propagator takes the
block diagonal form with $2 \times 2$ blocks
\begin{equation}
{\tilde D}^{-1}(\mu) = {1\over \lambda^2 (1-\mu^2) + \mu^2}
      \pmatrix{
         \mu (1-\lambda^2) & -\lambda \sqrt{1-\lambda^2} \cr
         \lambda \sqrt{1-\lambda^2} & \mu (1-\lambda^2)} ,
\end{equation}
and in topologically non-trivial background fields, the $|Q|$ additional
blocks are
\begin{equation}
\pmatrix{ \frac{1}{\mu} & 0 \cr 0 & 0 }
  \qquad {\rm or} \qquad
\pmatrix{ 0 & 0 \cr 0 & \frac{1}{\mu} }
\end{equation}
depending on the sign of $Q$.

The non-topological contribution
to the fermion bilinear, which is the part that will survive the infinite
volume limit, is
\begin{equation}  
\langle \bar \psi \psi \rangle = 
\langle \frac{1}{V} \sum_{\lambda > 0} \frac{2\mu (1-\lambda^2)}
 {\lambda^2 (1-\mu^2) + \mu^2} \rangle .
\label{eq:cond}
\end{equation}
The chiral condensate 
is dominated by the small, non-zero eigenvalues. In the thermodynamic
limit, it is given by the density of
eigenvalues at zero $\rho(0^+)$.

A physical quantity that is more sensitive at small quark masses
than the chiral condensate is the connected part of the scalar
susceptibility. The non-topological contribution to this quantity is
\begin{equation}
\chi_{a_0} = - {1\over V} \langle {\rm Tr}\tilde D^{-2} \rangle 
= \langle {d\over d\mu} \langle \bar\psi \psi \rangle_A \rangle
= \langle
 {1\over V} \sum_{\lambda > 0}
{2(1-\lambda^2) \left[ \lambda^2(1+\mu^2) - \mu^2 \right]
\over [\lambda^2(1-\mu^2) + \mu^2]^2} \rangle \quad .
\label{eq:chia0}
\end{equation}
This quantity is particularly sensitive to the $\mu$ dependence of the 
cancellations between the
$\lambda^2$ and $\mu^2$ terms in the second factor of the numerator.

Since chiral symmetry is exact on the lattice for the overlap Dirac
operator, the pion susceptibility is $\chi_\pi = {1\over\mu} <\bar\psi\psi>$
\cite{Robert},
and we can define a quantity
\begin{equation}
\omega  =  \chi_\pi - \chi_{a_0} 
= {1\over\mu} \langle \bar\psi \psi \rangle
-  \langle {d\over d\mu} \langle \bar\psi \psi \rangle_A \rangle
= {4 \over V} \Biggl\langle \sum_{\lambda > 0}
{\mu^2(1-\lambda_i^2)^2 
\over [\lambda_i^2(1-\mu^2) + \mu^2]^2} \Biggr\rangle\quad .
\label{eq:omega}
\end{equation}
This is a measure of the $U(1)_A$ symmetry breaking in a dynamical
theory.
It is strictly positive and is not sensitive to
cancellations like those within the sum for $\chi_{a_0}$ in
Eq.~(\ref{eq:chia0}). 
Due to the higher powers of the quark mass and eigenvalues
in the sum, it is more sensitive than the chiral
condensate to small eigenvalues. 

\section{Eigenvalue spectrum}

We will investigate the quark spectrum on the ensembles of pure SU(2) and
SU(3) gauge theory configurations listed in Table~\ref{tab:ensemble}.
At $N_T=4$, the critical coupling is $\beta_c=2.2986(6)$ for SU(2)
and $\beta_c=5.6925(2)$ for SU(3)~\cite{T_c}.
All ensembles in Table~\ref{tab:ensemble} are in the deconfined phase.

\begin{table}
\caption{ Pure gauge ensembles studied. All ensembles have an
extent of $N_T=4$ in the short direction. The number of configurations, the 
linear extent in
the spatial directions and the estimated temperature are also given.}
\label{tab:ensemble}
\vspace{2mm}
\begin{tabular}{|c|c|c|c|c|}\hline
Gauge group & $\beta$ & $N$ & $T/T_c$ & $L$ \\ \hline
SU(2) & $2.3$ & 50  & 1.0 & 8 \\ \hline
SU(2) & $2.4$ & 100  & 1.4 & 8 \\ \hline
SU(2) & $2.4$ & 200  & 1.4 & 16 \\ \hline
SU(2) & $2.5$ & 200  & 2.0 & 16 \\ \hline
SU(3) & $5.71$ & 200  & 1.03 & 16 \\ \hline
SU(3) & $5.75$ & 400  & 1.13 & 8 \\ \hline
SU(3) & $5.75$ & 200  & 1.13 & 12 \\ \hline
SU(3) & $5.75$ & 400  & 1.13 & 16 \\ \hline
SU(3) & $5.85$ & 400  & 1.38 & 8 \\ \hline
SU(3) & $5.85$ & 206  &  1.38 & 16 \\ \hline
\end{tabular}
\end{table}

\begin{figure}
\epsfxsize = 0.8\textwidth
\caption{
Histogram of the low lying eigenvalues of the staggered and overlap Dirac
operator for the SU(2) ensemble at $\beta=2.4$ and SU(3) ensemble
at $\beta=5.75$. Both lattices have a spatial volume of $16^3$.
For the overlap Dirac operator the exact zero modes are not shown.
}
\centerline{{\setlength{\epsfxsize}{7in}\epsfbox[-20 0 580 550]{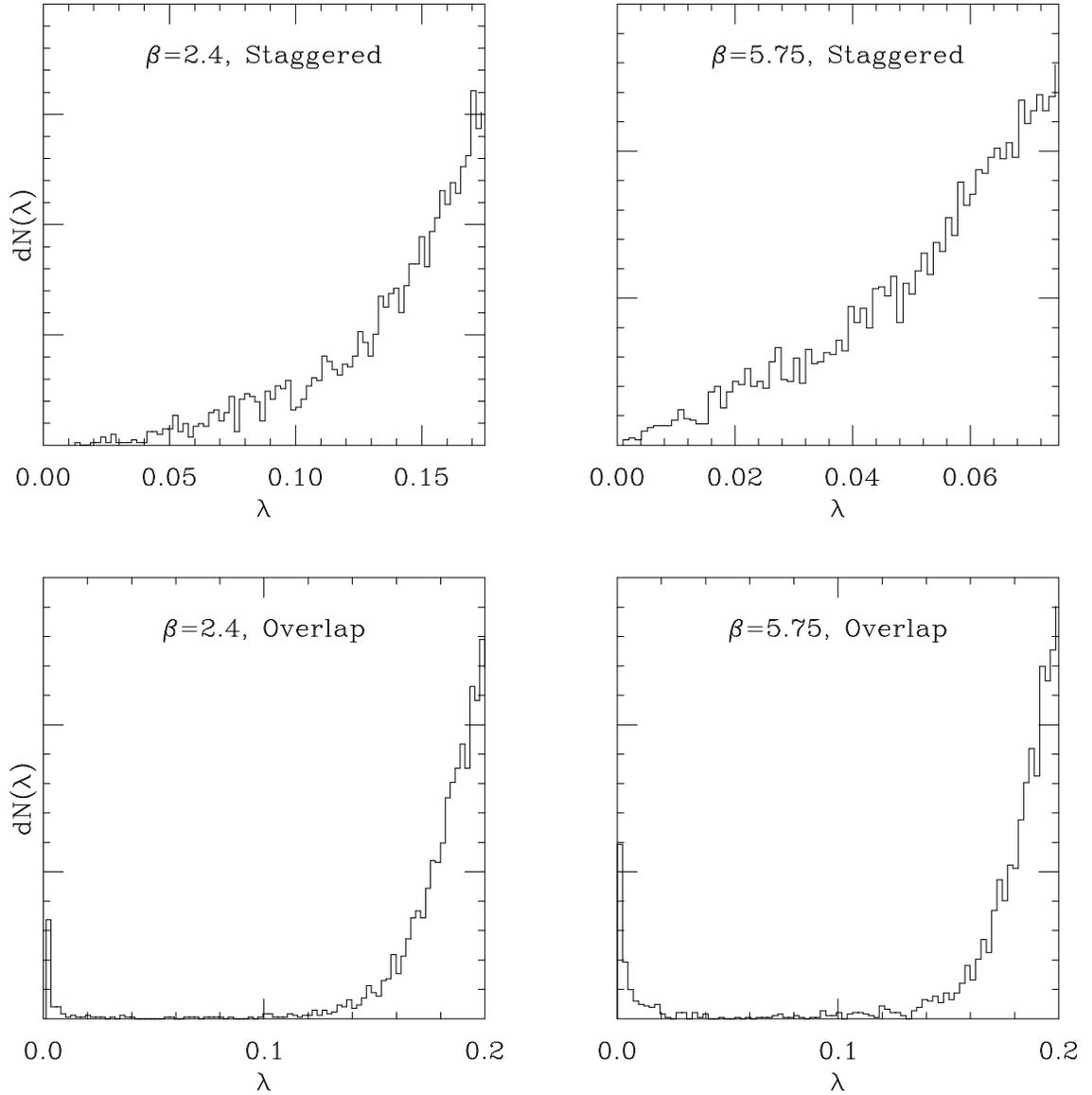}}}
\label{fig:histogram}
\end{figure}

The low lying eigenvalues of $H_0^2$ (including the exact zero modes) 
and $D_sD_s^\dagger$ ($D_s$ is
the massless staggered operator) were computed using the Ritz variational
technique~\cite{Ritz}. In order to deal numerically
with the overlap Dirac operator, it is necessary to use a representation
for $\epsilon(H_w)$, and we used the 
optimal rational approximation~\cite{EHN}.
We compare the low lying spectrum of the staggered and overlap Dirac
operator for two typical ensembles, SU(2) at $\beta=2.4$ and SU(3) at
$\beta=5.75$ both on $16^3\times 4$ lattices, in Fig.~\ref{fig:histogram}.
We see a remarkable difference. Both staggered and overlap spectrum
show a rapidly rising bulk. But while the staggered spectrum has a rapidly
decreasing tail of small eigenvalues, the overlap spectrum shows an
accumulation of very small, but non-zero, eigenvalues.

The spectral distributions of the overlap Dirac operator for all the
ensembles in Table~\ref{tab:ensemble} have the same qualitative features:
\begin{itemize}
\item There is a typical scale ($\lambda=0.05$) independent of the
ensemble that separates the spectrum into two parts: ``small'' ($< 0.05$)
and ``large'' ($> 0.05$) eigenvalues.\footnote{ We note
here that the smallest eigenvalue for the free overlap Dirac operator
with anti-periodic boundary conditions in the temperature direction
is $0.238$ and $0.200$ for a Wilson mass of $1.5$ and $1.8$ respectively
at $N_T=4$.}
\item Small eigenvalues occur in all topological sectors.
\item The number of small eigenvalues per unit lattice volume
remains roughly constant at fixed coupling under changes of the volume
and decreases as one goes deeper into the deconfined phase.
\end{itemize}

In this section, we focus on the statistical properties of the 
low-lying eigenvalues of the overlap Dirac operator. 
Our conclusion is that it is roughly consistent with the data to
associate each nonzero pair of small eigenvalues with an 
instanton---anti-instanton pair. Topics related to the magnitude of the
eigenvalues are discussed in the next section.

We approach the hypothesis in steps of increasing specificity. Let $n_+$ and
$n_-$ be the number of separated, localized instantons and anti-instantons
respectively. Then the topological charge of the configuration, as determined
by the fermions, is $Q = n_+-n_-$ and the total number of objects is 
$n \equiv n_+ + n_-$.

For infinite separation between instantons and anti-instantons,
this gives $n$ zero eigenvalues. At large but
finite separation, we expect $|Q|$ exactly zero eigenvalues and
$n-|Q|=2\ {\rm min}(n_+,n_-)$ small eigenvalues. Since the nonzero
eigenvalues come in $\pm \lambda$ pairs, we associate each nonzero
pair of small eigenvalues with an instanton---anti-instanton pair.  The
number of exact zero modes of $H_0^2$ in a fixed gauge field
background is equal to the net number of levels crossing zero in the spectral
flow of $H_w$~\cite{NN}. In addition, individual levels crossing zero
in the spectral flow of $H_w$ can be associated with instantons in the
background gauge field~\cite{EHN1,EHN2}. We performed a spectral flow
on several of the ensembles in Table~\ref{tab:ensemble} and found that
there is essentially a one-to-one agreement: If $n_+(n_-)$ levels
crossed zero from above (below), then there were 
$n-|Q|=2\ {\rm min}(n_+,n_-)$ 
small eigenvalues for the overlap Dirac operator in
addition to the rigorously expected $|Q|=|n_+-n_-|$ exact zero
eigenvalues.  Therefore our association of small eigenvalues of $H_0$
with instanton---anti-instanton pairs is well justified.

For finite temperature with $T>T_c$, the distribution of instanton sizes
increases steeply with instanton size until it is cutoff on the large size end
by $N_T$. Thus most of the instantons have a size not too far from $N_T$. We
will always use a simple model in which the instantons and anti-instantons are
assumed to be at sufficiently large separation that their interactions
due to the pure gauge action can be
neglected.

With the assumption that the $n$ objects are noninteracting, the number per
configuration should have a Poisson distribution
\begin{equation}
 P(n,\langle n \rangle) = \langle n \rangle^n {\rm e}^{-\langle n \rangle}/n!
\label{eq:poisson}
\end{equation}
where $\langle n \rangle$ is the average of the distribution. 
For a Poisson distribution,
the average and variance are equal. We present our data for the distribution
of $n$ as counted by the fermions in Table~\ref{tab:poisson}. 
The average of the distribution
per $8^3$ lattice volume is shown in the same table. The numbers in
parenthesis are the predictions of the distribution using the average value
in~(\ref{eq:poisson}).
We also write down the ratio of the variance and the average in the table.
The data are in fairly good agreement with the Poisson distribution.

To the extent that the $n$ zero and small eigenvalues can be associated with 
any sort of objects localized in 3-space or Euclidean 4-space, the number of 
objects and $n$ should, at fixed $N_T$ and fixed
coupling $g^2$, be proportional to the spatial volume $V$ in lattice units.
This is roughly supported by the SU(2) and SU(3) data in 
Table~\ref{tab:poisson}. 
If the localized objects are instantons, then we also expect the density to
decrease rapidly with decreasing $g^2$. This is also seen in the data. 

\begin{table}
\caption{Distribution of the number of topological objects along with
predictions from a Poisson distribution.}
\label{tab:poisson}
\vspace{2mm}
\begin{tabular}{|c|c|c|c|c|c|c|c|c|c|c|}\hline
        & SU(2) & SU(2) & SU(2) & SU(2) & SU(3)  & SU(3)  & SU(3)  & SU(3)  & SU(3)  & SU(3)  \\ \hline
$\beta$ & $2.3$ & $2.4$ & $2.4$ & $2.5$ & $5.71$ & $5.75$ & $5.75$ & $5.75$ & $5.85$ & $5.85$ \\ \hline
$L$     &  $8$  &  $8$  & $16$  & $16$  & $16$   & $8$   & $12$   & $16$   & $8$   & $16$   \\ \hline
$N$     & $50$  & $100$ & $200$ & $200$ & $200$  & $400$  & $200$ & $400$  & $400$  & $206$  \\ \hline
\hline
$n=0$ &11(10)&74(75)&27(28)&130(130)&1(1)&286(290)&83(77)&38(35)&379(378)&144(137)  \\ \hline
$n=1$ &13(16)&23(22)&57(55)&57(56)&5(6)&100(93)&63(73)&76(85)&19(22)&43(56)  \\ \hline
$n=2$ &14(13)&3(3)&58(54)&10(12)&8(16)&13(15)&39(35)&105(104)&2(1)&17(11)  \\ \hline
$n=3$ &8(7)&&32(35)&3(2)&28(27)&1(2)&11(11)&95(84)&&1(2)  \\ \hline
$n=4$ &2(3)&&16(17)&&46(35)&&3(3)&46(51)&&1(0)  \\ \hline
$n=5$ &2(1)&&4(7)&&38(35)&&1(1)&25(25)&&  \\ \hline
$n=6$ &&&3(2)&&34(30)&&&9(10)&&  \\ \hline
$n=7$ &&&3(1)&&17(21)&&&5(4)&&  \\ \hline
$n=8$ &&&&&10(14)&&&1(1)&&  \\ \hline
$n=9$ &&&&&6(8)&&&&&  \\ \hline
$n=10$ &&&&&2(4)&&&&&  \\ \hline
$n=11$ &&&&&3(2)&&&&&  \\ \hline
$n=12$ &&&&&2(1)&&&&&  \\ \hline
${\langle n \rangle/ V}$  &1.66&0.29&0.25&0.05&0.63&0.32&0.28&0.31&0.06&0.05 \\ \hline
$Ratio$ &0.97&1.09&0.93&0.99&1.15&1.08&0.92&1.02&0.90&0.83  \\ \hline 
\end{tabular}
\end{table}

At the next level of detail, we can distinguish between the contributions to
$n$ from $n_+$ and $n_-$. If the instantons and anti-instantons are thrown into
the configurations independently (without interaction), then for a fixed $n$
the relative probabilities in the $(n_-,n_+)$ distribution will be given by the
binomial coefficients
\begin{equation}
  B(n_+,n_-|n)={1\over 2^n} {n!\over n_+! n_-!}\quad .
\label{eq:binomial}
\end{equation}
Combining this with $\langle n \rangle$ values from the data gives predictions for the 
numbers $n_+$ and $n_-$ of positive and negative chirality states.
We combine $(n_+,n_-)$ with $(n_-,n_+)$ and compare the data and predictions
in Table~\ref{tab:binomial}.
In almost all of the cases, the difference is within the statistical error
expected from the value of the predicted number.

\begin{table}
\caption{Distribution of the number of instantons for a fixed number
of topological objects ( less than eight ) along with predictions from
Binomial distribution using the Poisson average.}
\label{tab:binomial}
\vspace{2mm}
\begin{tabular}{|c|c|c|c|c|c|c|c|c|c|c|}\hline
        & SU(2) & SU(2) & SU(2) & SU(2) & SU(3)  & SU(3)  & SU(3)  & SU(3)  & SU(3)  & SU(3)  \\ \hline
$\beta$ & $2.3$ & $2.4$ & $2.4$ & $2.5$ & $5.71$ & $5.75$ & $5.75$ & $5.75$ & $5.85$ & $5.85$ \\ \hline
$L$     &  $8$  &  $8$  & $16$  & $16$  & $16$   & $8$   & $12$   & $16$   & $8$   & $16$   \\ \hline
$N$     & $50$  & $200$ & $200$ & $200$ & $200$  & $400$  & $200$ & $400$  & $400$  & $206$  \\ \hline
\hline
$n=2,n_+=0$ &6(7)&2(2)&33(27)&6(6)&1(8)&6(8)&20(18)&54(52)&2(0)&8(6)  \\ \hline
$n=2,n_+=1$ &8(7)&1(2)&25(27)&4(6)&7(8)&7(8)&19(18)&51(52)&0(0)&9(6)  \\ \hline
\hline
$n=3,n_+=0$ &1(2)&&12(9)&0(0)&7(7)&0(0)&3(3)&25(21)&&0(0)  \\ \hline
$n=3,n_+=1$ &7(5)&&20(27)&3(1)&21(21)&1(1)&8(8)&70(63)&&1(1)  \\ \hline
\hline
$n=4,n_+=0$ &1(0)&&2(2)&&9(4)&&0(0)&9(6)&&0(0)  \\ \hline
$n=4,n_+=1$ &1(2)&&6(9)&&18(17)&&2(1)&30(26)&&1(0)  \\ \hline
$n=4,n_+=2$ &0(1)&&8(6)&&19(13)&&1(1)&7(19)&&0(0)  \\ \hline
\hline
$n=5,n_+=0$ &0(0)&&0(0)&&1(2)&&0(0)&2(2)&&  \\ \hline
$n=5,n_+=1$ &1(0)&&1(2)&&10(11)&&0(0)&9(8)&&  \\ \hline
$n=5,n_+=2$ &1(1)&&3(4)&&27(22)&&1(0)&14(16)&&  \\ \hline
\hline
$n=6,n_+=0$ &&&0(0)&&1(1)&&&0(0)&&  \\ \hline
$n=6,n_+=1$ &&&0(0)&&7(6)&&&3(2)&&  \\ \hline
$n=6,n_+=2$ &&&2(1)&&19(14)&&&2(5)&&  \\ \hline
$n=6,n_+=3$ &&&1(1)&&7(9)&&&4(3)&&  \\ \hline
\hline
$n=7,n_+=0$ &&&0(0)&&1(0)&&&0(0)&&  \\ \hline
$n=7,n_+=1$ &&&0(0)&&1(2)&&&0(0)&&  \\ \hline
$n=7,n_+=2$ &&&1(0)&&3(7)&&&1(1)&&  \\ \hline
$n=7,n_+=3$ &&&2(0)&&12(12)&&&4(2)&&  \\ \hline
\hline
\end{tabular}
\end{table}

Finally, the assumption of a non-interacting gas of instantons
and anti-instantons leads to a simple expression for the distribution
of topological charge,
\begin{equation}
T(Q)={\rm e}^{-\langle n \rangle}I_Q(\langle n \rangle)
\label{eq:top}
\end{equation}
where $I_Q$ is the modified Bessel function of order $Q$. Our data can
be cross-checked against this prediction also, and this is shown in
Table~\ref{tab:top}. Again, we see fairly good agreement with the
data.
We also note that $\langle Q^2 \rangle=\langle n \rangle$ 
under this assumption, and this is essentially the case when we
compare Table~\ref{tab:poisson} and Table~\ref{tab:top}.

This simple model of associating the zero and the small, non-zero eigenvalues 
with independent instantons and anti-instantons seems to account for the main
statistical properties of the $n_+,n_-$ data.

\begin{table}
\caption{ Distribution of topological charge along with the prediction based
on a non-interacting gas of instantons and anti-instantons using the average
from the Poisson distribution.}
\label{tab:top}
\vspace{2mm}
\begin{tabular}{|c|c|c|c|c|c|c|c|c|c|c|}\hline
        & SU(2) & SU(2) & SU(2) & SU(2) & SU(3)  & SU(3)  & SU(3)  & SU(3)  & SU(3)  & SU(3)  \\ \hline
$\beta$ & $2.3$ & $2.4$ & $2.4$ & $2.5$ & $5.71$ & $5.75$ & $5.75$ & $5.75$ & $5.85$ & $5.85$ \\ \hline
$L$     &  $8$  &  $8$  & $16$  & $16$  & $16$   & $8$   & $12$   & $16$   & $8$   & $16$   \\ \hline
$N$     & $50$  & $200$ & $200$ & $200$ & $200$  & $400$  & $200$ & $400$  & $400$  & $206$  \\ \hline
\hline
$Q=0$ &19(17)&75(76)&61(62)&134(136)&38(36)&293(297)&103(96)&101(110)&379(378)&153(143)  \\ \hline
$Q=1$ &21(22)&23(22)&82(86)&60(57)&67(65)&101(95)&72(82)&164(166)&19(22)&44(57)  \\ \hline
$Q=2$ &7(8)&2(2)&41(37)&6(6)&45(47)&6(8)&22(19)&86(83)&2(0)&9(6)  \\ \hline
$Q=3$ &2(2)&&14(11)&&26(28)&&3(3)&35(30)&&  \\ \hline
$Q=4$ &1(0)&&2(3)&&18(14)&&&12(9)&&  \\ \hline
$Q=5$ &&&&&3(6)&&&2(2)&&  \\ \hline
$Q=6$ &&&&&1(2)&&&&&  \\ \hline
$Q=7$ &&&&&1(1)&&&&&  \\ \hline
$Q=8$ &&&&&&&&&&  \\ \hline
$Q=9$ &&&&&&&&&&  \\ \hline
$Q=10$  &&&&&1(0)&&&&& \\ \hline
$\langle Q^2 \rangle/V$ &1.66&0.31&0.25&0.05&0.64&0.31&0.28&0.33&0.07&0.05  \\ \hline
\end{tabular}
\end{table}

\section { Small eigenvalues and a chiral condensate}

This section deals with the spectrum of the small, non-zero eigenvalues.
We begin with a discussion relating the spectrum to expectations from the
instanton---anti-instanton gas model. This is followed by an analysis of the
way in which these eigenvalues contribute to physical quantities such as
the chiral condensate.

In the previous section, we were only concerned with the number of
small eigenvalues (eigenvalues below $0.05$) in each configuration.
Our computation of the low lying eigenvalues of $H_o^2$ along
with the spectral flow of $H_w$ enables us to obtain this number
with very good certainty per configuration. Because of the numerical
accuracies involved, the small eigenvalues can be trusted only with
an absolute accuracy of $0.005$, and we can create a histogram of
the low lying eigenvalues with a bin width of $0.005$ giving
us a total of ten bins in each ensemble.

A detailed analysis of the relationship of small eigenvalues,
instantons, and the condensate at zero temperature can be found in 
\cite{Verbaarschot}.
It
was shown there that the spectrum from a model with fermions in a background of
dilute instantons and anti-instantons gives a chiral condensate and is
consistent with chiral random matrix theory.
Some discussion of the finite temperature spectrum appears in 
\cite{Schafer}.
We have used a much simpler version of these ideas.

The most important effect for widely separated instantons and anti-instantons 
is from the mixing of the $n$ would-be zero modes of the fermions. At infinite
separation, there would be $n$ modes with $\lambda=0$. At finite temperature
and in the continuum, the spatial tails of these wave functions are 
exponential. Further a configuration with {\em e.g.} $n=n_+$ has that number of
exact zero modes. This leads us to a model with instantons and anti-instantons
that are still distributed independently over the configurations and with an
$n \times n$ interaction matrix for the $n$ modes that is made up of 
$n_+ \times n_+$ and
$n_- \times n_-$ blocks on the diagonal and $n_- \times n_+$ and 
$n_+ \times n_-$ off-diagonal blocks. The diagonal blocks corresponding 
to instanton---instanton fermion mode interactions and to
anti-instanton---anti-instanton fermion mode interactions, including
self-interactions are zero. 
\begin{equation}
 T_{ij}= 0  \; \; \text {  if $i$ and $j$ are both instantons or both 
anti-instantons.}
\end{equation}
In the ``off-diagonal'' blocks there is an entry
for each ordered way that an instanton can be paired with an anti-instanton. 
It has the form
\begin{equation}
 T_{ij}=h_0 {\rm e}^{-d(i,j)/D} \; \; \text {  if $(i,j)$ is an 
instanton---anti-instanton pair.}
\end{equation}
The energy scale is determined by the constant $h_0$. The distance between the
instanton---anti-instanton pair is $d(i,j)$. The length scale of the
mode interactions is $D$. We expect $D$ to be of order $N_T$ since that is the
range of the continuum mode tails.

\begin{figure}
\caption{ Histogram of the small eigenvalues along with the
histogram from the toy model for $D=1,2,3$ with the
appropriate value of $\langle n \rangle$ from Table~\ref{tab:poisson} 
for the SU(2) ensemble at $\beta=2.4$ and SU(3) ensemble
at $\beta=5.75$. Both lattices have a spatial volume of $16^3$.
}
\centerline{{\setlength{\epsfxsize}{7in}\epsfbox[-20 180 592 460]{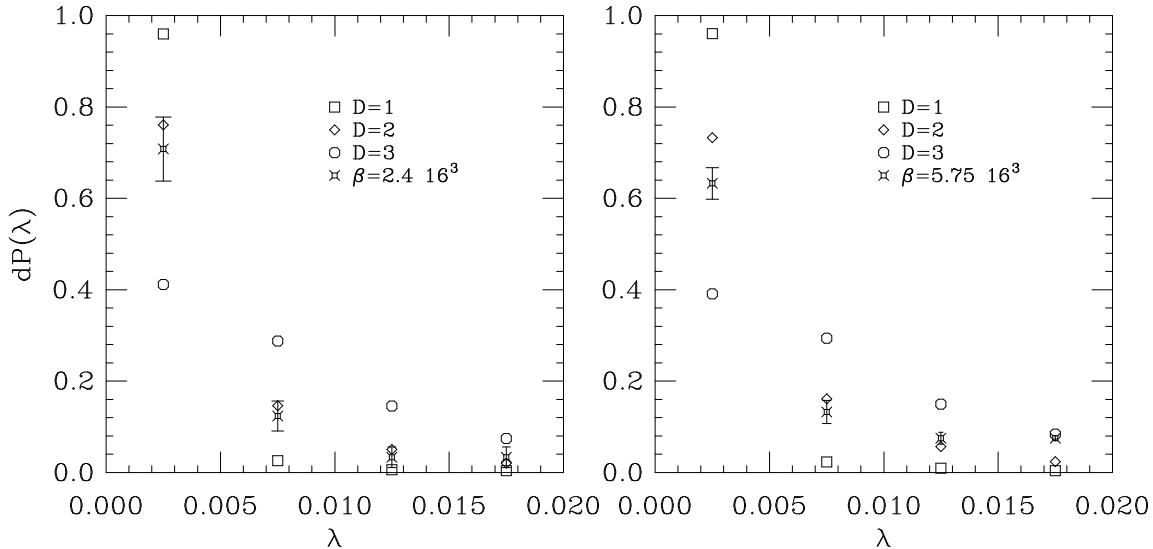}}}
\label{fig:toy}
\end{figure}

The next step is to generate data from the model and compare it with the real
data. The
model calculation is in the continuum with periodic boundary conditions, so
that $d(i,j)$ is the shortest distance connecting the pair.
The $(n_+,n_-)$ values are 
distributed according to Eqn.~(\ref{eq:poisson}) and 
Eqn.~(\ref{eq:binomial}) with $\langle n \rangle$ obtained from
Table~\ref{tab:poisson}. The positions of 
the instantons and anti-instantons are selected at random. Then the matrix
elements of $T$ are computed. Finally the eigenvalues of $T$ are computed. Note
that for a matrix with the form of $T$, the spectrum has $|Q|$ zeros and 
$(n-|Q|)/2$ pairs $\pm \lambda$. This process is repeated for many 
configurations. The resulting set of eigenvalue distributions is then
compared with the data from the various ensembles. The conclusion is that 
a value of $D=2$ best
fits the data for all the ensembles. We have shown the comparison between
the ``real'' data and the toy model for two of the cases in Fig.~\ref{fig:toy}.
Thus the spectrum of the small eigenvalues can also be explained by the simple
instanton---anti-instanton gas picture.

A physical role for these modes is suggested by the fact that the chiral
condensate is determined by the infinite-volume fermion spectral density at
zero eigenvalue.
The most straightforward thing to do would be to use the computed eigenvalues
to do the sums in Eqns.~(\ref{eq:cond}) -- (\ref{eq:omega})
for the condensate, $\omega$, and $\chi_{a_0}$. 
However, we need the $\mu \rightarrow 0$ limit, and 
it is apparent from the forms of the sums that they are very sensitive 
to the positions of the eigenvalues when $\mu$ is very small. 
As mentioned earlier in this section, the small eigenvalues are not
precisely known, so that the value of the sum is unreliable  
for $\mu$ less than about 0.01, and this is not small enough to be useful.

\begin{figure}
\caption{The stochastic estimates of $\langle \bar\psi\psi \rangle$,
$\omega$ and $\chi_{a_0}$ for the SU(2) ensemble at $\beta=2.4$ and
SU(3) ensemble at $\beta=5.75$. Both lattices have a spatial volume of
$16^3$.}
\centerline{{\setlength{\epsfxsize}{7in}\epsfbox[-30 0 600 770]{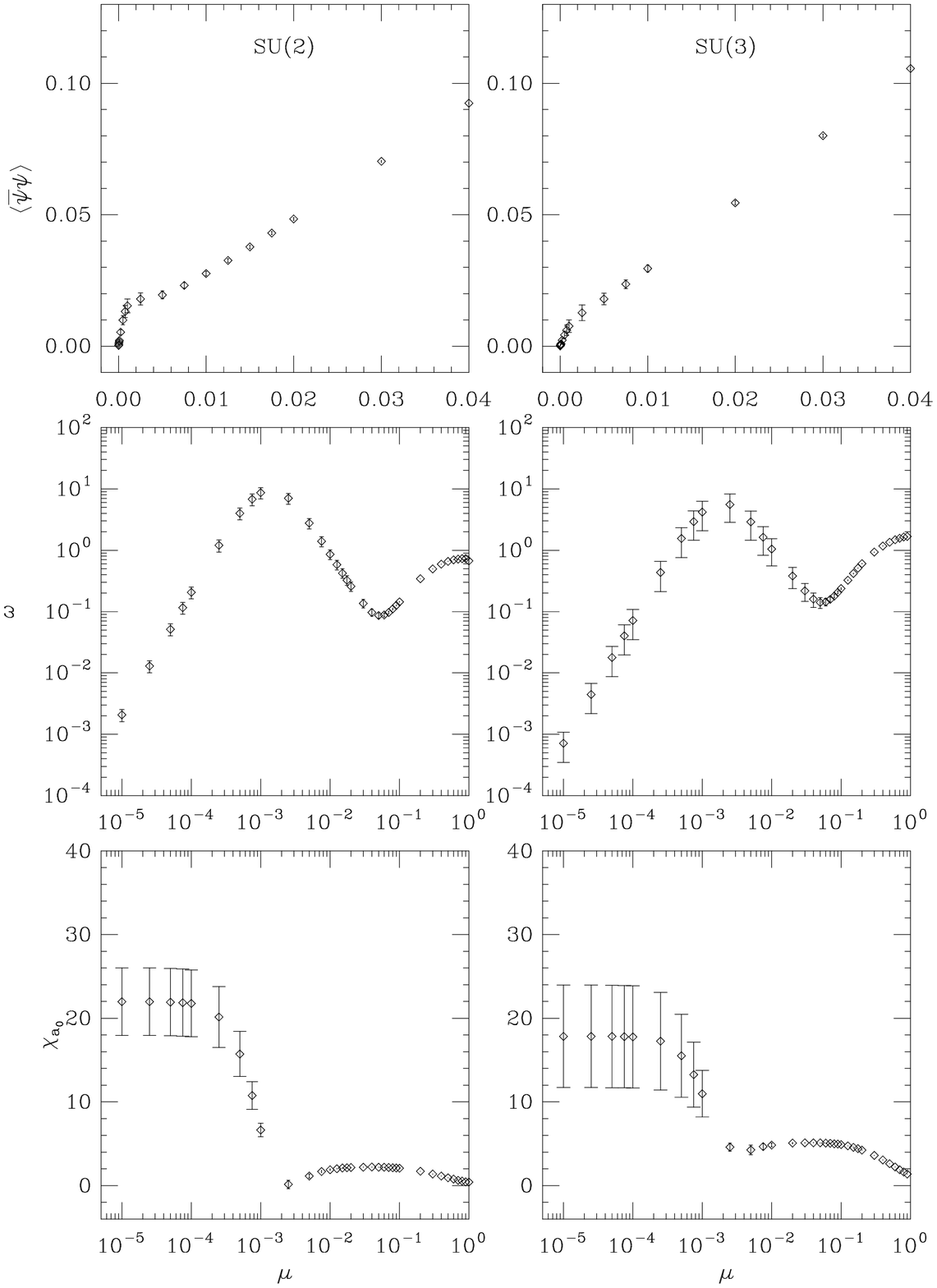}}}
\label{fig:pbpa0}
\end{figure}

However, it is possible to compute the chiral condensate, 
$\omega$, and 
$\chi_{a_0}$ by a stochastic method using the overlap Dirac operator.
The method is specific to the overlap Dirac operator and is described in
Ref.~\cite{Robert,EHN2}. Using this method, we have computed
the condensate, the anomalous U(1)$_A$ breaking, and the scalar
susceptibility for the SU(2) ensemble
with $\beta=2.4$ on the $16^3$ lattice and for the SU(3) ensemble
with $\beta=5.75$ also on the $16^3$ lattice. The results are
shown in Fig.~\ref{fig:pbpa0}.
If the steep decrease of the condensate toward zero below $\mu=0.002$ is 
interpreted as a finite volume effect, then these data are evidence for chiral
symmetry breaking. Additional evidence for chiral symmetry breaking can
be seen in the plots for $\omega$.
We see there is a quark mass region in the
non-topological contribution to $\omega$
from $\mu= 0.002$ to $0.05$ where the expected divergence of
${1\over\mu}$ sets in.
The region below $\mu=0.002$ is where the
finite volume effects become large.

The shapes of the curves in Fig.~\ref{fig:pbpa0} are consistent with the
qualitative features of the computed eigenvalues: small eigenvalues
concentrated close to $\lambda=0$, a sparsely populated region around
$\lambda=0.05$, and the dense eigenvalues of  the bulk of the spectrum
beginning at larger eigenvalue. 
The strongest effect is a dip in $\chi_{a_0}$ due to the presence of
two terms in the sum in Eqn.~(\ref{eq:chia0}) with opposite signs
along with a spectral distribution that has two regions (a small
eigenvalues region and a bulk region) separated by a sparsely
populated region around $\lambda=0.05$ as seen in Fig.~\ref{fig:histogram}.

One may ask whether the simple model, which
reproduces the properties of the small eigenvalues, gives a chiral condensate.
The short answer to that question is ``Yes". Our model is just a simplification
of the model in Ref.\cite{Verbaarschot}, which does give a condensate.
The longer answer comes after testing the simplified model itself for the
necessary properties.
With $\langle n \rangle/V$ fixed and
increasing $V$, the staircase function 
\begin{equation}
    N(\lambda,V) = \int_0^{\lambda} d\sigma \rho(\sigma,V)
\end{equation}
in the large $V$ and small $\lambda$ region needs to approach a
function of $\lambda V$ that is linear for large $\lambda V$.
$N(\lambda,V)$ as computed with modest statistics from the model eigenvalues 
is consistent with that behavior. 
Thus it appears that the eigenvalues in the ``small region'' of the
spectrum are sufficient to give rise to a condensate.

\section { Conclusions}

We have studied the spectrum of the overlap Dirac operator on several
quenched ensembles in the deconfined phase. We find clear evidence for
topology in the deconfined phase. We also found that
the topological susceptibility
decreases sharply as we go deeper into the deconfined phase. The gauge
field contains instanton and anti-instanton like objects
which are well described as a dilute gas. This is supported by
a part of the spectrum of the overlap Dirac operator consisting of
``small'' (in our case $< 0.05$) eigenvalues well separated from the bulk.
This small part is consistently described as arising
from a dilute gas of instantons and anti-instantons. 
It appears that the small eigenvalues produced by such a dilute gas
are sufficient to create a chiral condensate and hence spontaneous
chiral symmetry breaking even in the deconfined phase of quenched gauge
theories.

\acknowledgements We would like to thank the Aspen Center for Physics
where part of this work was carried out. We would also like to
thank Herbert Neuberger, Philippe de Forcrand and Poul Damgaard
for discussions. UMH and RGE were supported in part by DOE contracts
DE-FG05-85ER250000 and DE-FG05-96ER40979. RGE was also supported by DOE
contract DE-AC05-84ER40150 under which the Southeastern Universities
Research Association (SURA) operates the Thomas Jefferson National
Accelerator Facility.
Computations were performed on the CM-2, QCDSP, and workstation
cluster at SCRI and UC Davis.

\newpage

\end{document}